\documentclass[twocolumn]{article}
\newif\ifpdf
\ifx\pdfoutput\undefined
  \pdffalse
\else
  \pdfoutput=1
  \pdftrue
\fi

\ifpdf
  \usepackage[pdftex]{epsfig}
\else
  \usepackage{epsfig}
\fi
\begin{document}

\title{Ordering And Partial Ordering In Holmium Titanate And
 Related Systems}
\author{R Siddharthan$^1$\thanks{Presently at the Laboratoire
de Physique Th\'{e}orique, Ecole Normale Sup\'{e}rieure,
24 rue Lhomond, 75231 Paris Cedex 05, France, and at the
Laboratoire de Physique des Solides, Universit\'{e} Paris-Sud,
Bat.~510, 91405 Orsay, France.} , B S
Shastry$^{1,2}$ and
A P  Ramirez$^2$ \\ \em$^1$ Department of
 Physics, Indian Institute of Science, Bangalore 560012, India \\
\em $^2$ Bell Laboratories, Lucent Technologies,
600 Mountain Avenue, Murray Hill, NJ 07974, USA}

\maketitle

\begin{abstract}
We take another look at two compounds which have been discussed as
possible realizations of ``spin ice'', namely holmium titanate and
dysprosium titanate.  As we have earlier observed, holmium titanate
does not display ice-like behaviour at low temperatures because
the long ranged dipolar interactions between spins are strong compared
to the nearest neighbour interactions.  We show, analytically, that
the true ground state of this system must be fully ordered, but
simulations only reach partially ordered states because there are
infinite energy barriers separating these from the true ground state.
We also show that the true ground state of our model of dysprosium
titanate is also fully ordered, and offer some explanations as to why
simulations and experiments show ice-like behaviour.  We discuss
the effect on these systems of an applied magnetic field.  Finally, we
discuss several other models which show similar partial or full
ordering in their ground states, including the well known Ising model
on the fcc lattice.
\end{abstract}

\section{Introduction}
\label{intro}

With the wide interest in the physics of disorder and frustrated
magnetism, pyrochlore magnets have attracted great attention in recent
years \cite{ramirez94}, and it is especially interesting to consider
pyrochlores well approximated by the Ising model. We recently
discovered \cite{nature} that dysprosium titanate, an Ising
pyrochlore, exhibits a ground state entropy very much like that of
ice. Anderson had long ago predicted that this should happen for a
nearest-neighbour Ising pyrochlore.  However, the story is not quite
so simple here: the dominant interaction is really a long-ranged
dipole-dipole interaction.  Moreover, the  similar Ising
pyrochlore holmium titanate has very different low-temperature
behaviour.  We had explored the reasons for this in our earlier papers
\cite{nature,prl}; here we take those arguments further. 

We observed earlier \cite{prl} that simulations of a model of the
Ising pyrochlore holmium titanate suggest that it has a partially ordered
ground state, which is degenerate but has no entropy per particle (the
degeneracy being of the order $2^L$ where $L$ is the system length,
rather than $2^{L^3}$), and there appears to be a first order phase
transition from the high temperature paramagnetic phase to the above
partially ordered phase. In this article we clarify the nature of this
partial ordering, and explain it without recourse to simulations. 
Actually, we
show that the ordering of the true ground state here is in fact
complete, but there are numerous low-lying partially ordered
metastable states which are separated from the true ground state by
infinite energy barriers. It is easy for the simulation (and,
presumably, the real compound) to get stuck in one of these states on
cooling, and impossible to reach the true ground state in finite time
thenceforth. Moreover, this is also the true ground state of our model
of dysprosium titanate; but both the model and the experiments
\cite{nature} suggest
a finite ground state entropy characteristic of nearest-neighbour 
``spin ice''. This, we suggest, is because this system has
stronger nearest-neighbour interactions and settles into an
ice-like state at a higher temperature (over $1$~K), and cannot then
be easily dislodged from this into the true ground state. Our exact
results, and also our simulations, support and substantiate
our original suggestions of a transition to partial ordering, in contrast 
to recent suggestions
to the contrary \cite{gingras}, namely that that our model for
holmium titanate should have an ice-like ground state. 

We also substantiate the major results and the underlying model from
our earlier work \cite{nature}, namely that
there is a low temperature entropy observed in dysprosium titanate, it
is decreased in the presence of a magnetic field, and the interactions
in the system are dipole-dipole magnetic interactions and an isotropic
superexchange.  The fact that our simulations with a magnetic field
reproduce experimental results quite well, qualitatively and
quantitatively, confirms our calculation of the dipole moment (which
is the only thing that couples with the field) and our estimate of the
superexchange.  Moreover, the ground state in the presence of a strong
field is not the same as the zero-field ground state.  While the
experiments were done on powdered samples, simulations suggest that
the behaviour of dysprosium titanate in a magnetic field is very
direction dependent. A strongly direction dependent ordering was 
initially suggested on the basis of the specific heat measurement done
on a powdered sample \cite{nature}. 
We compare the experimental data on powder
samples with simulations averaged over large numbers of directions,
compare sharp features in both, and make important predictions for
possible future experiments on single crystals.

Next, we use the insight from the ground state analysis to
progressively reduce the pyrochlore Ising model to a sequence of
simpler models which display similar behaviour: namely, a six-vertex
model on the ``diamond lattice''  with non-local interactions, which
reproduces all the essential behaviour of holmium titanate, and has a
fully ordered ground state and several metastable partially ordered
low-lying states (section~\ref{sec:vertex}; a six-state magnetic model
on the fcc lattice, which actually has the sort of partially ordered
ground state that the simulations had suggested for holmium titanate,
and also reproduces the important behaviour of holmium titanate
(section~\ref{sec:sixstate}); and the well-known Ising model on the
fcc lattice, which has been studied before \cite{isingfccrefs} and is
known to have exactly the same sort of partial ground state ordering
that concerns us here (section~\ref{sec:fccising}).  Along the way, we
also introduce a square-lattice vertex model, by analogy with the
above diamond-lattice model, which may be worthwhile to study in its
own right.

\begin{figure}[ht]
\begin{center}
 \ifpdf
  \epsfig{file=tri.pdf, width=7cm, clip=}
 \else
  \epsfig{file=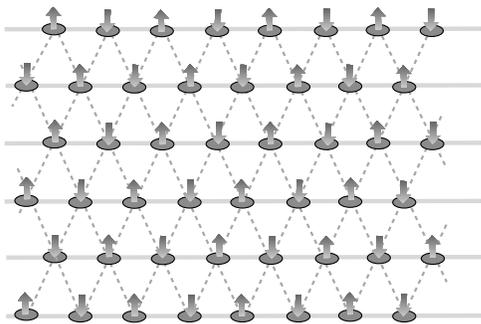, width=7cm, clip=}
 \fi
\end{center}
\caption{\label{tri} Anisotropic triangular lattice, with antiferromagnetic
 Ising interaction $J$ along solid lines and $J' < J$ along dashed lines.
 In ground state, each chain along solid lines is ordered, but there is no
 order along dashed lines.}
\end{figure}

Actually, the simplest example of partial ordering in an Ising system
is perhaps the triangular lattice antiferromagnet, with interaction
$J_1$ along bonds in one direction (say parallel to the $x$ axis),
$J_2 < J_1$ along bonds in other directions (Figure \ref{tri}), and
$J_1, J_2$ positive (antiferromagnetic).  Then we have perfect
antiferromagnetic order along a line of sites in the $x$ direction,
but each atom on an adjacent chain is frustrated so the adjacent chain
(which also has perfectly antiferromagnetic order) has two possible
configurations with respect to the first, and the system as a whole
has a degeneracy $2^L$ where $L$ is the number of chains. This system
is exactly solvable \cite{wannier}, and has no finite-temperature
phase transition.  This situation is quite relevant to what happens in
our system.  The analogous model in three dimensions, the fcc Ising
model, is discussed at the end of the paper and in the cited
references.

\section{Ordering in our model of Ho$_2$Ti$_2$O$_7$}
\label{sec:holmium}

\begin{figure}[ht]
\begin{center}
  \ifpdf
  \epsfig{file=pyro.pdf, width=6cm,clip=}
  \else
  \epsfig{file=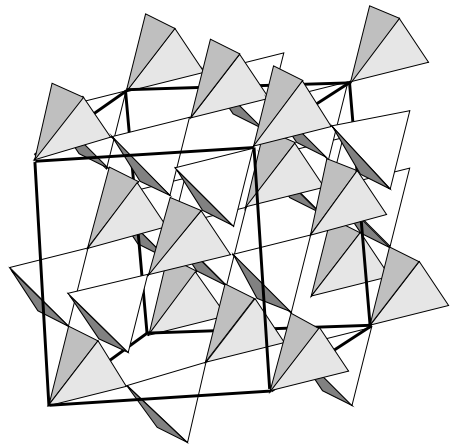, width=6cm,clip=}
  \fi
\end{center}
\caption{\label{pyro} The conventional unit cell of the pyrochlore,
 a lattice of corner-sharing tetrahedra with fcc symmetry.
}
\end{figure}

\begin{figure}[ht]
\begin{center}
  \ifpdf
  \epsfig{file=latbas.pdf, width=6cm,clip=}
  \else
  \epsfig{file=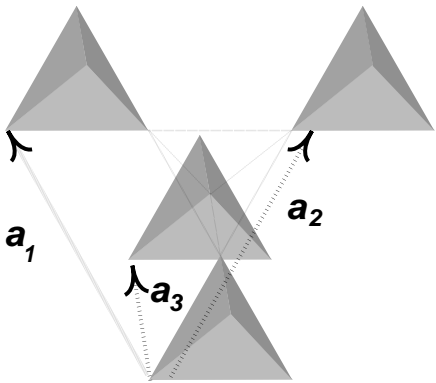, width=6cm,clip=}
  \fi
\end{center}
\caption{\label{latbas} The pyrochlore lattice can be
generated by translating a tetrahedron along vectors $\mbox{\bf a}_1$,
$\mbox{\bf a}_2$, $\mbox{\bf a}_3$. Tetrahedra of the opposite
orientation are formed from the corners of these (thin black lines).}
\end{figure}

We briefly recapitulate our model of holmium titanate. The underlying
lattice is the pyrochlore lattice, a lattice of corner-sharing
tetrahedra of two possible orientations,  which is well visualised as
an fcc lattice of tetrahedra (Figure~\ref{pyro}). 
It may be generated by taking a single
tetrahedron of one orientation and translating it by the primitive
basis vectors of the fcc lattice (Figure~\ref{latbas}); 
the tetrahedra of the other orientation
emerge automatically by this procedure---see Figure~\ref{latbas}. 
Thus we use the lattice
vectors
\begin{eqnarray}
{\bf a}_1 & = &  (r, \sqrt 3 r, 0 ), \nonumber \\
{\bf a}_2 & = &  \left( -r, \sqrt 3 r, 0 \right), \nonumber \\
{\bf a}_3 & = &  (0, 2r/\sqrt 3, -2r \sqrt{2/3}). \label{latticevec}
\end{eqnarray}
with a basis of atoms located at
\begin{eqnarray}
 {\bf x}_0 & = & (0,0,0) \nonumber \\
 {\bf x}_1 & = & 
            \left( r, 0, 0 \right), \nonumber \\
 {\bf x}_2 & = & 
          \left( \frac{1}{2} r, \frac{\sqrt{3}}{2} r, 0 \right), \nonumber \\
 {\bf x}_3 & = &  
          \left( \frac{1}{2} r, \frac{1}{2\sqrt{3}} r, \sqrt{\frac{2}{3}} r
              \right) \label{basis}
\end{eqnarray}
In this system $r$ is around 3.54 {\AA}ngstroms. 

We have Ising spins (the
$f$~electron states of the holmium atoms) located at these points. The
local Ising axis is the line joining the centres of the two adjoining
tetrahedra: thus, each spin points directly out of one tetrahedron, and
into the next one. The spins carry magnetic moments corresponding
to $J=8$, $g_s$ (the Land\'e factor) $= 1.25$. Based on this, the
expected nearest-neighbour dipole-dipole interaction has an energy of
$\pm 2.35$~K, the sign depending on the alignment of the spins:  one
pointing out, one in is preferred. However, the experimental compound 
(unlike its cousin, dysprosium titanate) has properties which we can only
explain by postulating a significant nearest-neighbour superexchange,
which we estimate from high temperature expansions for the
susceptibility to be around $1.9$~K with opposite sign to the dipolar
interaction. So, with the convention that
an Ising spin $S=1$ if it points out of an ``up'' tetrahedron and
$S=-1$ otherwise, we write the Hamiltonian as follows:
\begin{eqnarray}
H & = &  \sum_{i,j} J_{ij} S_i S_j  \\
J_{ij} & = & 0.45 \mbox{~Kelvin, for nearest neighbour spins} \\
J_{ij} & = & \frac{\mu_0}{4\pi} g_s^2 \mu_b^2 J^2 \left[
       \frac{{\bf n}_i\cdot{\bf n}_j}{r_{ij}^3} -
        3 \frac{({\bf n}_i \cdot {\bf r}_{ij})({\bf n}_j \cdot {\bf r}_{ij})}
          {r_{ij}^5} \right], \nonumber \\ 
  & &\mbox{~further neighbours}
\end{eqnarray}
where ${\bf n}_i$ is a unit vector pointing along the Ising
axis at site $i$ in the outward direction from an ``up'' tetrahedron.
Thus, this system has a
drastically reduced nearest-neighbour interaction energy, which means
the importance of the further-neighbour interactions increases. We saw
that with only a long ranged dipole-dipole interaction between these
Ising spins, the behaviour seems to change little from the
nearest-neighbour Ising model which has a finite ground state entropy;
we speculate that the substantially different behaviour of
holmium titanate is due to the significantly greater importance of
further-neighbour interactions. As noted earlier, simulations of
such a model do predict a freezing of the system at around $0.7$~K,
in agreement with experiment.

\begin{figure}[ht]
\begin{center}
 \ifpdf
   \epsfig{file=threetet.pdf, width=7cm, clip=}
 \else
   \epsfig{file=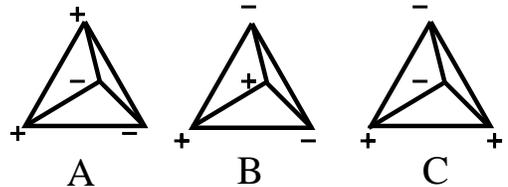, width=7cm, clip=}
 \fi
\end{center}
\caption{\label{threetet} Three possible ground state configurations
$A$,$B$, $C$ of a tetrahedron. We will denote by $A'$, $B'$, $C'$ these
respective configurations with all spins reversed. `$+$' indicates a spin
pointing out of the tetrahedron, `$-$' a spin pointing into the
tetrahedron.}
\end{figure}

We now find the ground state (GS) for this system. For clarity we
refer to the two different orientations of tetrahedra that
occur as ``up'' and ``down'': each ``up'' tetrahedron
shares corners with only ``down'' tetrahedra, and vice versa.
Consider a single tetrahedron: it has six possible ground
states, each of which has two spins pointing out and two in.
We label these states $A$, $A'$, $B$, $B'$, $C$, $C'$ where
$A'$ is $A$ with the directions of all spins reversed, and
similarly $B'$ and $C'$ (Figure~\ref{threetet}).
Consider a cluster of sites consisting of a single ``up''
tetrahedron and its translation by the three primitive
lattice vectors of the fcc lattice (as in Figure~\ref{latbas}), 
and long-ranged dipole-dipole
interactions spanning the entire cluster plus a nearest-neighbour
superexchange, as described above. We find the
GS of this ``tetrahedron of tetrahedra'' by the
unobjectionable method of enumerating each of the $2^{16}$ allowed
states on a computer, and picking out the lowest-energy ones.

\begin{figure}[ht]
\begin{center}
 \ifpdf
  \epsfig{file=fourtet.pdf, width=8cm, clip=}
 \else
  \epsfig{file=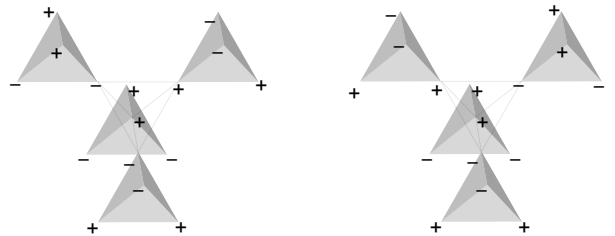, width=8cm, clip=}
 \fi
\end{center}
\caption{\label{fourtet} 
  Two of the twelve allowed ground states for a cluster of
  four tetrahedra. Each of the twelve states is related to the
  others by rotations or reflections. This consideration by
  itself leads to partial ordering in the ground state of
  the full system, as described in the text.
  }
\end{figure}

We find that the GS of this cluster is twelve-fold
degenerate. Two are shown in Figure~\ref{fourtet}. All others
are rotations/reflections of these (which are
reflections of each other).  In each GS, as one might
expect, only states satisfying the ice rule occur.  Furthermore, only
two kinds of ice-like configurations occur: $A$ and $A'$, or $B$ and
$B'$, or $C$ and $C'$. They occur twice each. In the case shown in the
figure, once the configuration of tetrahedron at the origin is fixed
as $C$, the tetrahedron at ${\bf a}_3$ must have the opposite
configuration ($C'$); and the tetrahedron at ${\bf a}_2$ may have
either configuration ($C$ or $C'$), but tetrahedron ${\bf a}_1$ must
have the opposite configuration to ${\bf a}_2$.  With the 
previously quoted values for
the dipole and superexchange interactions,
the configurations in
Figure~\ref{fourtet} have an energy $-7.5$~K, and the next lower
energy configurations have an energy $-6.9$~K.

If we note that each tetrahedron in its ground state has a dipole
moment, which is perpendicular to two possible lattice translation
vectors, the rule is this: in every GS only configuration of
two opposite kinds occur (say $C$ and $C'$), which therefore have
antiparallel magnetic moments, and two tetrahedra separated by a
vector perpendicular to these moments must have opposite
configurations. These are the twelve ground states allowed for this
cluster, but we are making no theoretical argument for this;
this is what we learn from brute-force enumeration of
states. In any of these states, moreover, the included ``down''
tetrahedron also has an ice-ruled configuration.

Now consider the entire fcc lattice of tetrahedra (that
is, the pyrochlore lattice). Once we fix the configuration
of a single ``up'' tetrahedron (say, as $C$) each of its neighbouring
tetrahedra, with which it forms part of a cluster of the above sort,
must have the same or opposite configuration (say $C$ or $C'$); and by
extending further from each of these, this must be true for the entire
lattice. We can also see that if we travel in either of the two
lattice directions perpendicular to the magnetic moment of these
tetrahedron configurations, we must have a perfectly alternating
sequence (say $C$, $C'$, $C$, $C'$, \ldots ); but when travelling in a third 
direction we have no ordering rule. So we get a ground state ordering in two
directions but not in the third, and a degeneracy exponential in
$L$ (the system length) rather than $L^3$ (the system volume), 
precisely as the simulations suggested. 

This is not the whole story, though. The system can be equally well
described in terms of ``down'' tetrahedra, so the same sort of
ordering should be evident if we describe a ground state 
configuration using ``down'' tetrahedra. So only two configurations
for these tetrahedra should be allowed, each of which is the other
with all spins reversed. But we note immediately that the two ``down''
tetrahedra displayed as grey lines in the two clusters in
Figure~\ref{fourtet} do not have opposite
configurations. It follows that the two cluster configurations in
that figure cannot both occur in the ground state: only one can. {\em This
immediately implies that the ordering sequence in a third direction
is not random, but constant} (say, $C$, $C$, $C$, \ldots ). 

So the true ground state for our model of holmium titanate 
is only twelvefold degenerate, and viewed in terms of configurations
of either upward or downward tetrahedra, consists of alternating
ordering of opposing configurations in two directions but a constant
configuration in a third direction. However, the partially ordered 
states are also very low
in energy, and moreover a system stuck in such a state can only
get out by flipping entire planes of tetrahedron configurations,
which is impossible in the thermodynamic limit. So simulations
tend to get stuck in such states and in other ``domainised'' states
and the chances of a given simulation actually hitting a true
ground state are very small---unless we use some sort of specialised
``cluster'' algorithm which may not imitate the dynamics of the real
system very well.

This is exact except for one thing: we have ignored interactions
between further-neighbour tetrahedra, so effectively confined our
interaction range to the 5th neighbour, which is the maximum separation
of spins in two adjacent tetrahedra.  As long as only nearest-neighbour
``up'' tetrahedra interact, so that the whole system really can be
decomposed into clusters as in Figure~\ref{fourtet}, our picture is totally
correct. In the presence of long ranged interactions, luckily the
nature of the system (with four different local $z$ directions) is
such that the effects of the  more distant tetrahedra cancel heavily and have
little effect on the energy of a single cluster. Thus, if one
take a particular ``up'' tetrahedron in the ground state, its energy
turns out to be $-21$~K because of interactions with the immediately
neighbouring ``up'' tetrahedra (to which the cluster argument applies),
but only $-0.12$~K because of interactions with all other tetrahedra
in the system. The effect is not only small, but tends to stabilise
this order. (With a random ice-like configuration, this energy averages
to zero, but fluctuates considerably from site to site.)  
Recall furthermore that the cost of disturbing a single 
four-tetrahedron cluster from its ground state (Figure~\ref{fourtet}) 
is at least $0.6$~K, and in fact much more since each tetrahedron
is shared by four such clusters.  

\begin{figure}[ht]
\begin{center}
 \ifpdf
  \epsfig{file=spcutoff.pdf, width=8cm, clip=}
 \else
  \epsfig{file=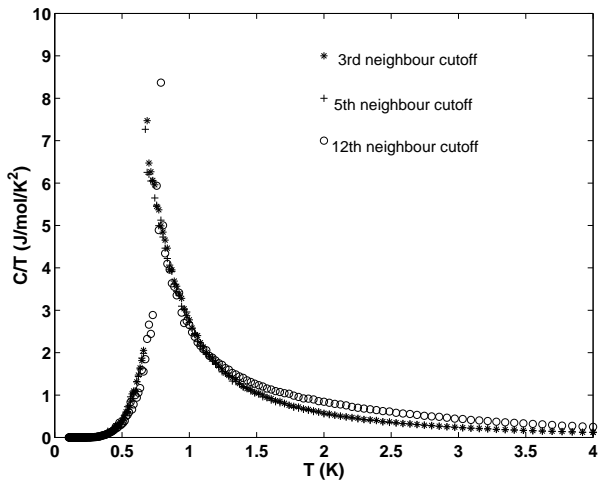, width=8cm, clip=}
 \fi
\end{center}
\caption{\label{spcutoff}
   The simulated specific heat when the interaction is cut off
   at the third ($R \le 2r$), fifth ($R \le 2.646 r$), and
   twelfth ($R \le 4r$) nearest neighbour distances ($r =
   3.54$~\AA\ roughly). The position
   of the phase transition and the plot of the specific heat near
   the transition hardly change at all on increasing the range of
   the interaction; but one needs longer
   equilibriation times with increased interaction ranges. The
   simulations show a significant energy drop at the transition,
   suggesting that it is first order.
  }
\end{figure}

So we can be satisfied that the long ranged interactions will not
affect the above conclusions.  In fact, the simulations too don't show
much dependence on the range of the interaction, provided it extends
to at least the third neighbour (Figure~\ref{spcutoff}), as indeed we
had argued in our earlier paper, where we cut off the interaction at
the fifth-neighbour distance \cite{prl}.  The suggestion
\cite{gingras} that ice-like behaviour is restored by cutting off
after the 10th neighbour seems untenable to us, and we do not observe
it in our simulations even on extending the interaction to the 12th
neighbour (which is halfway across our sample).  Possibly the
different results  obtained in \cite{gingras} is
due to additional approximations
involved, such as Ewald sums, and a somewhat
smaller sample size; our simulations use no approximations in
calculating the energy except the cutoff, which as we have seen, is
quite justifiable. Longer simulations on bigger systems  may
throw more light on this question, but we now turn to some other
interesting aspects of the problem.

This true ground state ordering is more easily visualised (though less easily
analysed) with the cubic unit cell rather than our parallelepiped; 
this is discussed in section~\ref{sec:sixstate}.

\section{Dysprosium titanate}

\begin{figure}[ht]
\begin{center}
 \ifpdf
  \epsfig{file=nndy.pdf, width=8.5cm, clip=}
 \else
  \epsfig{file=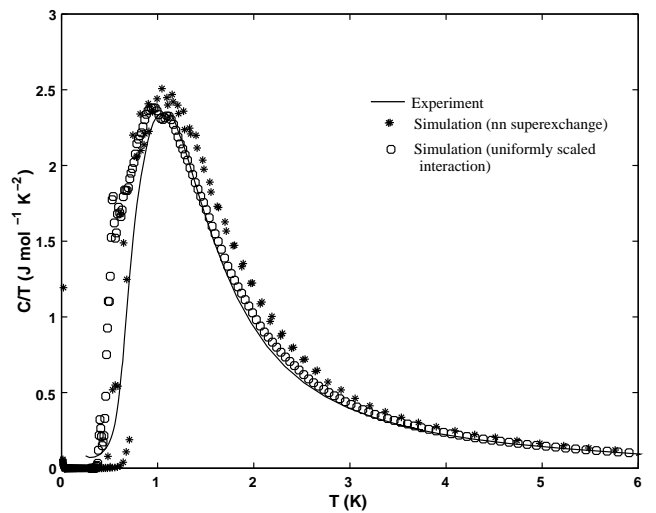, width=8.5cm, clip=}
 \fi
\end{center}
\caption{\label{nndy}
  A comparison with experimental data for dysprosium titanate
  of a simulation using nearest-neighbour superexchange and
  long-ranged dipole-dipole interactions, and a simulation using
  uniformly scaled down dipole-dipole interactions (as was done
  in ref.~\cite{nature}).
 }
\end{figure}

Dysprosium titanate, which we earlier reported as showing ice-like
behaviour experimentally and in simulations, appears to have a much
weaker superexchange between nearest neighbours.  With a
nearest-neighbour-only model of the superexchange, we find we need a
superexchange of around $+ 1.1$~K, that is, the nearest-neighbour
interaction is around $-1.25$~K compared to the bare dipole-dipole
value of $-2.35$~K.  With these numbers, we get a reasonable agreement
of the simulation with experiment (Figure~\ref{nndy}).  However, we
get even better agreement by uniformly scaling down the long-ranged
dipole-dipole interaction by roughly this factor. (This is what was
reported in our earlier paper.) This suggests that the superexchange
is not strictly nearest-neighbour but extends over to further
neighbours.

We need to understand why Ho$_2$Ti$_2$O$_7$-like
behaviour is not observed in this case.
With either of the two models above (small nearest-neighbour-only
superexchange, or uniformly scaled-down dipole-dipole interaction)
the ground state for the cluster of four tetrahedra remains the same
as before, the difference in energy from the next-lowest state too
remains roughly the same, and the above arguments should still go
through.  

\begin{figure}[ht]
\begin{center}
 \ifpdf
  \epsfig{file=phasetrans.pdf, width=9cm, clip=}
 \else
  \epsfig{file=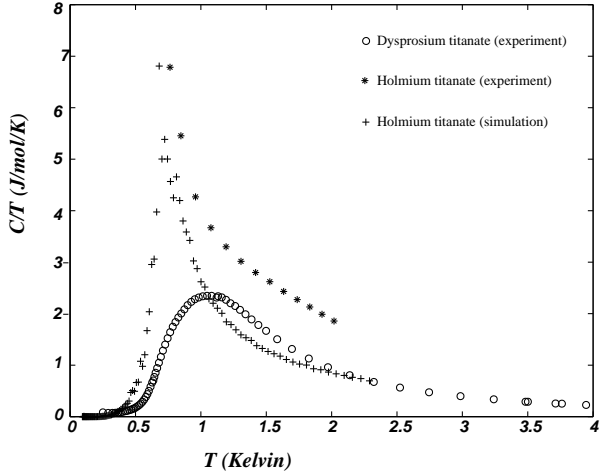, width=9cm, clip=}
 \fi
\end{center}
\caption{\label{phasetrans}
  Specific heat curves of dysprosium titanate (experimental, and
  simulations give excellent agreement) and holmium titanate
  (experimental, and a simulation retaining dipole-dipole
  interactions to 12 nearest-neighbour distances). 
  By the time the ordering temperature is reached (which is
  expected to be the same for both systems) dysprosium titanate
  is already frozen into an ice-like configuration from which it
  would find it hard to locate a lower-energy state.
  }
\end{figure}

The difference is that, because of the stronger nearest-neighbour
interactions here, this system undergoes a crossover from a
paramagnetic phase to an ice-ruled phase at a considerably higher
temperature (greater than $1$~K); and by the time it cools down 
to the temperature ($< 0.7$~K) where we
expect a transition of the sort described here, it is already stuck
in a disordered ice-like state and (because of the stronger
nearest-neighbour interactions) cannot easily break out of this state
to access other states. It appears that the temperature of the
crossover to the ice-like phase is dictated by the nearest-neighbour
interactions. In dysprosium titanate these have an energy of around
$1.3$~K, and hence the ice rule is already in place by the time we go down to
$0.7$~K and the spins are almost frozen, thus the ordering transition no
longer has a chance to occur. In holmium titanate, on the other hand, the
nearest-neighbour interaction is around $0.4$~K, thus at $0.7$~K the
system is in no sense frozen,  plenty of   spin
flips take place
 and the ordering transition occurs (Figure~\ref{phasetrans}).

So while our arguments show that the true ground state here is ordered
and only twelve-fold degenerate, the system tends to get stuck in
fairly generic ice-like states. We have checked that the energy of the
disordered low-temperature state of the system in our simulations is
always slightly but significantly higher (by around 0.5\% to 1\%) 
than the energy of the fully
ordered state if we include the full long-ranged interactions in
calculating the energy.

\begin{figure}[ht]
\begin{center}
 \ifpdf
  \epsfig{file=integb.pdf, width=8.5cm, clip=}
 \else
  \epsfig{file=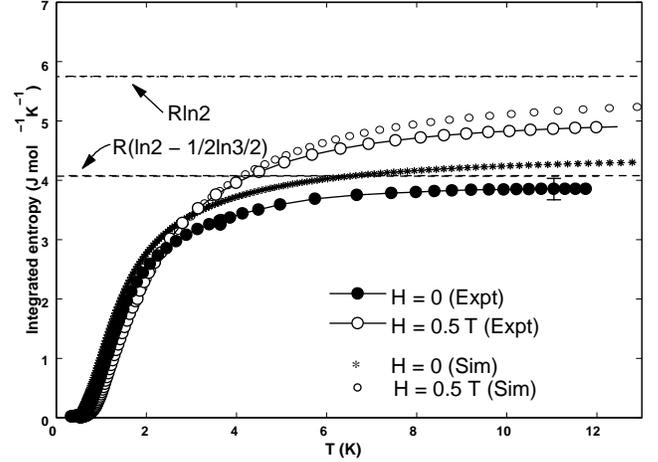, width=8.5cm, clip=}
 \fi
\end{center}
\caption{\label{integb}
  The integrated specific heat per unit temperature in simulations
  without a magnetic field and with a half-tesla magnetic field.
  This shows the entropy gained over the ground state entropy.  The
  entropy is expected to be $R \log 2$ (dotted line) at high
  temperatures;  the integrated value falls short of this, indicating
  a ground state entropy, but the ground state entropy is reduced in
  the presence of a magnetic field.  Both the experimental data, taken
  from our earlier paper~\cite{nature}, and the simulation results are
  plotted here for easy comparison.  Based on simulations, we
  suggest that a magnetic field of
  around 3~Tesla should recover all or nearly all the ground state
  entropy.  
  }
\end{figure}

  The low-temperature states we see are
governed mainly by the ice rule (though evidence of some local
ordering of 4-clusters can be seen) and are probably macroscopic in
number. 
This is why we observe an anomaly in the integrated entropy which we
earlier attributed to a possible ground state entropy. 
This low temperature ``entropy'' (as estimated from
Figure~\ref{integb}) 
is around 10\% lower than $1/2 \log (3/2)$, which is itself an
underestimate by around 10\%. 
Very much the
same thing is likely to be true in the real system (dysprosium
titanate) too: its true ground state is ordered but the system can
almost never access this ground state.  The measured ground state
entropy here is closer to $1/2 \log (3/2)$, in fact a bit more;
it is probably a bit less than the true ground state entropy of
nearest-neighbour spin ice, though.

In the presence of a magnetic field, some interesting things happen.
As reported earlier \cite{nature}, some of the observed ground state
entropy is recovered experimentally; we see this also in simulations
(Figure~\ref{integb}).  Since only the dipole moment couples to the
field, the quantitative agreement in this curve is an additional
confirmation of our model of co-existing dipole-dipole interactions
and superexchange.
The curves are similar in features and the amount
of entropy recovered is also roughly the same.  In stronger fields,
sharp spike-like features start to show up in the experimental
specific heat curves at low temperatures.  Here, too, we find
reasonable qualitative and quantitative agreement between simulations
and experiment.  All this confirms that our calculation of the dipole
moment of the $f$~electrons and our supposition that the reduced
energy scales are due to another interaction (superexchange) are
correct, since the interaction with a magnetic field is purely
magnetostatic.
The experiments were done using powder samples, and the simulations
show that the behaviour is strongly dependent on the direction of
the field.  To compare with powder averaged experimental results,
we would need a very large number of simulations in random directions,
which we have not done to our satisfaction.

The nature of the ground state also depends on the field direction,
and with a sufficiently strong field and a suitable field direction
the ground state may not even satisfy the ice rule.  For instance,
with a field along the $z$ axis in Figure~\ref{latbas} (which
corresponds to the $[1 1 1]$ direction with the conventional unit cell)
the ground state consists of three spins pointing into each upward
tetrahedron and one pointing out of it.

\begin{figure}[ht]
\begin{center}
 \ifpdf
  \epsfig{file=bfielddirs.pdf, width=8.5cm, clip=}
 \else
  \epsfig{file=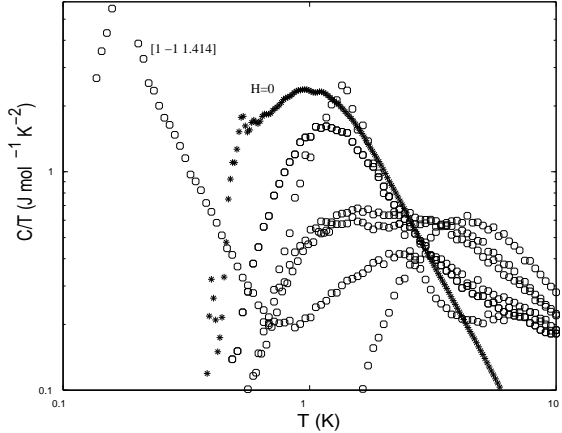, width=8.5cm, clip=}
 \fi
\end{center}
\caption{\label{bfielddirs}
   Specific heat in the presence of a $2$~Tesla magnetic field,
   for various directions of the field. 
 }
\end{figure}

\begin{figure}[ht]
\begin{center}
 \ifpdf
  \epsfig{file=mvsb.pdf, width=8cm, clip=}
 \else
  \epsfig{file=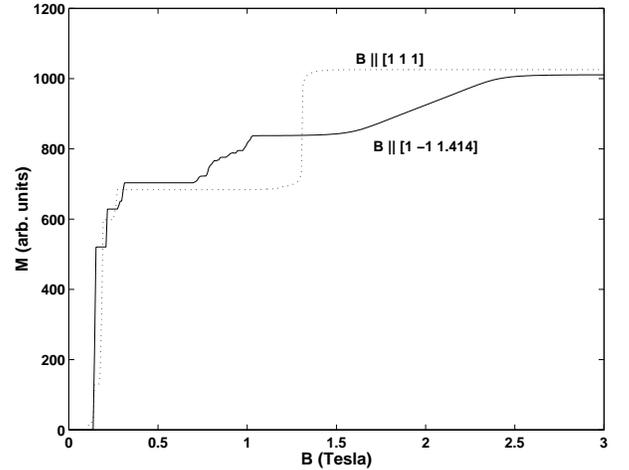, width=8cm, clip=}
 \fi
\end{center}
\caption{\label{mvsb}
  The growth of magnetisation in the simulation sample, for
  magnetic field in the $[1 \overline{1} \sqrt{2}]$ and $[1 1 1]$ directions, 
  at a temperature of $0.2$~K.
 }
\end{figure}

It is interesting to ask how the transition to a different ground state
occurs as one slowly turns on a magnetic field at low temperatures.
Figure~\ref{mvsb} shows the result of doing this in a simulation at
$0.2$~Kelvin, for a field in the $[1 \overline{1} \sqrt{2}]$ and $[1 1 1]$
direction.  The system seems to go through several magnetic
transitions before reaching its fully polarised state.

All these features would be averaged over in the experiments on
the powder samples, and single crystals of these materials could
turn out to be worth studying in their own right.

\section{An analogous six-vertex model}
\label{sec:vertex}

If we examine the nature of states just above the
transition in simulations of our holmium titanate model, we find
that a large fraction of the tetrahedra are already in the ice-ruled state.
This suggests that the sharp transition here is not the
transition from paramagnetism to an ice-like phase,
but a phase transition from
a disordered ice-like state to an ordered or partially ordered
state.

To check that this is the case, we can try putting the ice constraint
in by hand, and check that the phase transition is still reproduced
at the same temperature. The model we get then is a form of
six-vertex model on the ``diamond'' lattice, whose sites
are the centres of the tetrahedra in the pyrochlore lattice. 
The six-vertex model has been widely studied on the square lattice;
the diamond lattice, like the square lattice, has a coordination number
of four and can be divided in two sublattices, but is three-dimensional.
So we study a system where one assigns arrows to bonds on the
diamond lattice, such that each site (or ``vertex'') has two arrows 
pointing in at it and two pointing away: so six kinds of vertices are
possible. But unlike conventional six-vertex models, we don't assign
different weights to these six vertices, since all of them are really
equivalent here; instead, the thermodynamics comes from interactions
between different vertices. In other words, we have a Hamiltonian
of the sort
\begin{equation}
 H = \sum_{i,j} \equiv \sum_{i,j} J(c(i),c(j), {\bf r}_i - {\bf r}_j),
\end{equation}
where $c_i$ is the configuration (a six-valued variable) of the $i$-th
vertex, and $J$ is the interaction energy of vertices $i$ and $j$,
which depends not only on their configurations but on the vector
joining them (thanks to the underlying direction-dependent dipole-dipole
interaction). We have to calculate the pairwise $J$'s appropriately.

What we do is the following: we note that the sites on the diamond
lattice fall into two sublattices, corresponding to up and down
tetrahedra. First consider adjacent
vertices (adjacent corner-sharing tetrahedra). The internal
interactions between these spins can be separated into nearest-neighbour
interactions, which we can assume has already been taken
into account via the ice rule, and next-neighbour interactions, which
we can equally include by considering only next-neighbour vertices (that
is, nearest-neighbour tetrahedra of like orientation). 
We thus ignore interactions between nearest-neighbour vertices and
consider interactions only between
next-neighbour vertices, or nearest-neighbour vertices on a single
sublattice. The interaction between two such vertices is the energy
of interaction between the two corresponding tetrahedra, as given by
the sum of interaction energies of all pairs of spins. We ignore all
further-range interactions. But the interactions already included,
if carried out over all vertices over both sublattices, will actually
double-count the pairwise spin-spin interactions: we therefore also
insert a factor of half.

Formally, we can write
\begin{equation}
\label{vertexham}
H = \sum_{\{i,j\} } E_{ij} \equiv \sum_{\{ i,j \} } 
         E({\bf S}_i, {\bf S}_j, {\bf r}_i - {\bf r}_j)
\end{equation}
and then break this sum up into nearest-neighbour terms, next-neighbour
terms, and so on; throw out the nearest-neighbour terms because we have
used them in enforcing the ice constraint; and group the next few
terms into pairs where one spin belongs to one ``up'' tetrahedron, the
other spin belongs to an adjacent ``up'' tetrahedron. 
\begin{eqnarray}
H & = &  \sum_{ \{ \alpha, \beta \} } \sum_{m=1}^4 \sum_{n=1}^4
      E({\bf S}_{\alpha m}, {\bf S}_{\beta n}, {\bf R}_\alpha - {\bf R}_\beta
              + {\bf x}_m -{\bf x}_n) \nonumber \\
  & &+ \mbox{other terms}
\end{eqnarray}
where the sum is over neighbouring ``up'' tetrahedra at sites
$\alpha$ and $\beta$, and we sum the interaction of each of the
four spins at $\alpha$ with each of the four spins at $\beta$. 
${\bf R}_\alpha$ is the position of spin $1$ on tetrahedron $\alpha$,
likewise ${\bf R}_\beta$, and $x_m$ are as in equation~(\ref{basis}).
All the ``other terms'' involve spins separated by three times the
nearest-neighbour distance or more, so we drop them. We can
do precisely the same grouping of terms for the ``down'' tetrahedra;
we do both, and insert a factor of half. Thus our
Ising spin Hamiltonian is reduced to the vertex Hamiltonian~(\ref{vertexham})
with appropriately chosen interaction energies $J$, extending only
to the nearest neighbour on the same sublattice. 

As before, we simulate this Hamiltonian. First, a word on how we do this.
Flipping a single bond will not do: it will destroy the ice constraint
on both adjoining vertices. We must find a closed loop, a set of bonds
whose arrows lead from vertex to vertex and return to the starting
vertex, and flip the whole loop at one go. Such ``loop
algorithms'' have been discussed previously \cite{evertz} and
it has been pointed out that, in a six-vertex model, every line of
arrows if followed must return to the starting vertex and every configuration
is accessible via loop-flips alone, so a random loop-flip algorithm is
ergodic. The earlier algorithms have several improvements and
optimizations; however, they are concerned with conventional
vertex models where different vertices have different weights but
don't interact, and
cannot be completely translated to this situation. We found it
sufficient to merely pick up
random starting sites, form loops randomly, calculate the energy
difference, and flip them according to the Metropolis algorithm.

\begin{figure}[ht]
\begin{center}
 \ifpdf
  \epsfig{file=spcomp.pdf, width=8.5cm, clip=}
 \else
  \epsfig{file=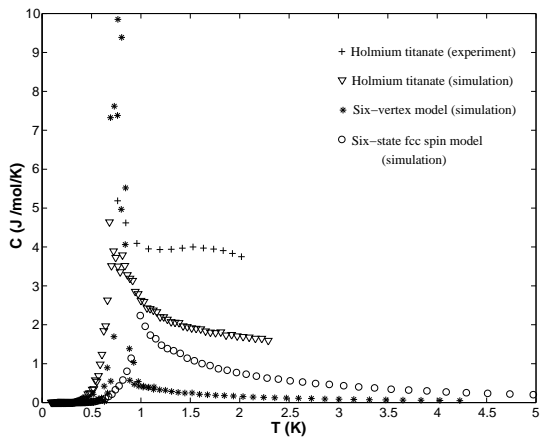, width=8.5cm, clip=}
 \fi
\end{center}
\caption{\label{spcomp}
  Comparison of specific heat curves for holmium titanate (the real
  system); our model of holmium titanate; the six-vertex model of
  section~\ref{sec:vertex}; and the six-state spin model of
  section~\ref{sec:sixstate}
  }
\end{figure}

The results are shown in Figure~\ref{spcomp}.
It exhibits a phase transition at exactly the point where both the
real system, and our model for it, do. This is a distinctly first order
phase transition. Thus we have verified that the phenomenon driving
this phase transition is not the formation of ice-like tetrahedra, but
the further ordering of tetrahedra that have already attained
ice-rule configurations; and we have displayed a fairly simple 
vertex model which shows the same features as our Ising pyrochlore.

The ground state of this system would be expected to be fully ordered,
but typically only partially ordered states are accessible. The
argument is similar to that in the case of holmium titanate, and
the simulations bear this out.

\section{A square lattice vertex model}

\begin{figure}[ht]
\begin{center}
 \ifpdf
  \epsfig{file=squareice.pdf, width=5.5cm, clip=}
 \else
  \epsfig{file=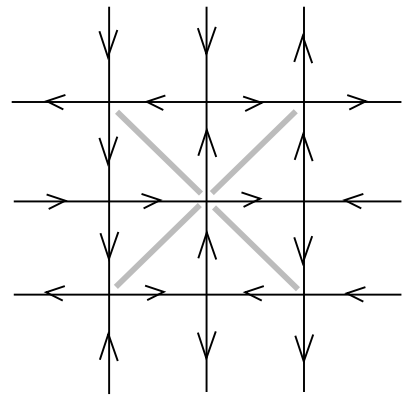, width=5.5cm, clip=}
 \fi
\end{center}
\caption{\label{squareice} 
  An ice model on the square lattice, with similar properties
  to the earlier, diamond-lattice model.  Interactions are
  between diagonally-opposite vertices (grey lines).
 }
\end{figure}

The sort of physics involved can perhaps be better seen in a
square-lattice vertex model.  Such models have been extensively
studied \cite{baxter}, but the thermodynamics has typically arisen
from assigning different weights to different vertices;  instead we
give the same weight to all vertices, but consider interactions
between diagonally-opposite vertices (the grey lines in
Figure~\ref{squareice}).  We ignore nearest-neighbour interactions for
the same reason as earlier, ie that is taken care of by assigning an
ice rule. 

\begin{figure}[ht]
\begin{center}
 \ifpdf
  \epsfig{file=sqrweights.pdf, width=7.5cm, clip=}
 \else
  \epsfig{file=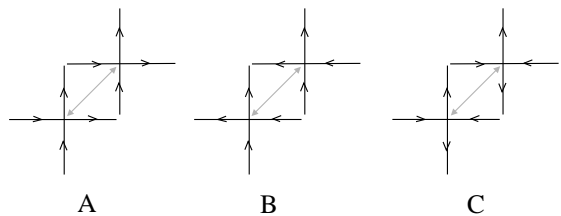, width=7.5cm, clip=}
 \fi
\end{center}
\caption{\label{sqrweights} 
  The three possible interactions between neighbouring
  vertices.  All other possibilities are symmetry related,
  or zero.  In particular, reversing all arrows on a single
  vertex will simply change the sign of the interaction.
 }
\end{figure}

\begin{figure}[ht]
\begin{center}
 \ifpdf
  \epsfig{file=sqricegs.pdf, width=7.5cm, clip=}
 \else
  \epsfig{file=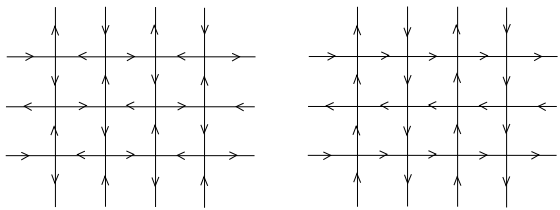, width=7.5cm, clip=}
 \fi
\end{center}
\caption{\label{sqricegs} 
  With interaction energies $A$, $B$, $C$ between vertices calculated
  from dipole-dipole interactions between spins aligned along their
  edges, we obtain a ground state as on the left.  But a slightly
  different choice of weights will yield a ground state as on the
  right, and there is the possibility of a ``level crossing'' between
  the ground states for a choice $A-B=2C$.
 }
\end{figure}

The possible interactions are shown in Figure~\ref{sqrweights};  for
symmetry reasons, we need have only three interaction parameters,
all other non-zero interactions can be obtained from these by
rotation, reflection, or inversion of one or both vertices (inverting
a single vertex will simply change the sign of the interaction energy).
If we calculate the interaction parameters from an assumed
dipole-dipole interaction between magnets aligned along the edges
connecting the respective vertices, we obtain $A=-8.6678$,
$B=10.753$, $C=-10.345$ (arbitrary units).  The ground state
then looks as shown on the left of Figure~\ref{sqricegs}; but a very
slight alteration in the choice of $A$, $B$ and $C$ will give a ground
state as shown on the right of Figure~\ref{sqricegs}, and the choice
$A-B=2C$ may lead to rather interesting results.  The model is
probably worth studying both in its own right and because of the
long historical interest square-lattice vertex models have held; but
since it is not really related to the rest of this article, we
postpone further discussion of it to a future work.

\section{Multistate spin model on an fcc lattice}
\label{sec:sixstate}

In our vertex model earlier, we had two sublattices, and interactions
only within a single sublattice. Apart from the ice-rule constraint,
the two sublattice could just as well be noninteracting. So the
next logical step is to separate the two sublattices. We consider
an fcc lattice, with a six-valued variable at each site. Only
nearest-neighbour interactions are considered, and as before, the value
of the interaction is determined from the underlying pyrochlore 
Ising variables.
The major difference with the vertex model case is that we have now
forgotten about the ``down'' tetrahedra: an arbitrary configuration
of four neighbouring ``up'' tetrahedra would not satisfy the
ice rule for the enclosed ``down'' tetrahedron, but we are no longer
worrying about that now.

It turns out that the dynamics of interaction between these ``up''
tetrahedra takes care of that for us. The system displays a phase
transition at very nearly the same temperature as the vertex model
and the Ising pyrochlore (Figure~\ref{spcomp}), 
and at temperatures just above the transition
the configuration is such that nearly all the ``down'' tetrahedra in
the correspondingly-configured pyrochlore
would satisfy the ice rule; at zero temperature the system is partially
ordered, in exactly the way we observed in holmium titanate, but the
partially ordered states in this case really are the ground states.

The partial ordering is now more easily visualised with the conventional
cubic unit cell rather than the parallelepiped which we used. Note
first that each ice-ruled state of a tetrahedron has a dipole moment,
perpendicular to the side connecting the two inward-pointing spins 
and to the side connecting the two outward-pointing spins. If one looks at
the cubic unit cell, we can see that the six allowed values of this dipole
moment are along the three edges of this cube: so what we have is
a six-state magnetic model on an fcc lattice where each spin can point
along one of the Cartesian axes. In the ground state, one of these
axes is picked out, so that each spin points along the same line in
one of two opposite directions; each plane perpendicular to this
direction is antiferromagnetically ordered; and perpendicular to this
plane, the ordering is random. 

It is tempting to use the total dipole moments of
the tetrahedra as the site variables, and for the interaction simply to 
use their mutual magnetic interactions,
since we know that the dipole-dipole
interaction favours antiferromagnetic ordering in planes perpendicular
to the spins; but it turns out that this is not the ground state
of such a system. Only by using the actual interaction energies of the
tetrahedra do we obtain such a ground state.
However, there is an obvious connection between
this system and a well studied-problem, which we turn to in the next
section.

\section{Ising model on the fcc lattice}
\label{sec:fccising}

The nearest-neighbour antiferromagnetic Ising model on the 
fcc geometry has been studied by several
authors, and its ground state is known to have exactly the sort of ordering
we are considering.

\begin{figure}[ht]
\begin{center}
 \ifpdf
  \epsfig{file=fcc.pdf, width=7cm, clip=}
 \else
  \epsfig{file=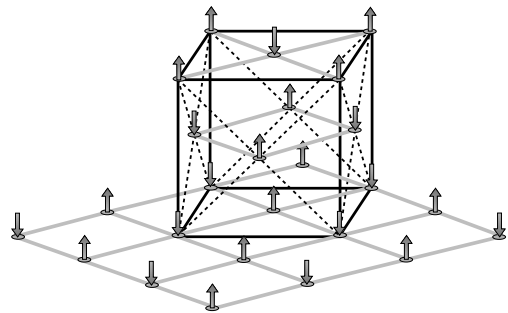, width=7cm, clip=}
 \fi
\end{center}
\caption{\label{fccising} 
  Ising antiferromagnet on an fcc lattice, with nearest
 neighbour interaction $J$ (solid grey
 lines) in the plane and $J' <J$ (dotted lines) between planes.
 Each plane orders antiferromagnetically but they stack up in
 a random manner. Surprisingly, this is also the ground state
 when $J'=J$.
 }
\end{figure}

The ordering is easy to understand if there is a bit of
anisotropy in the system: consider Figure~\ref{fccising}, where
we have an fcc crystal, and within the $x$--$y$ plane and planes
parallel to it there is an antiferromagnetic interaction $J$ between
nearest neighbours, but out of the plane there is an interaction
$J' < J$. Then the planes prefer to order antiferromagnetically,
but adjacent planes have zero interaction energy regardless of their
relative ordering, so the ordering along the $z$ axis is random. 
The interesting thing is that this remains true even when $J'=J$:
the only change is a new factor of $3$ in the degeneracy because
of the new rotational symmetry of the system.

The order of the transition is also of interest. In the isotropic case, it is
a first order transition. With strong anisotropy (a weak inter-plane
coupling), however, we would expect a second-order transition because
the system effectively is like weakly interacting 2D Ising systems,
which have a second order transition at the Onsager temperature.
In fact, simulations suggest that for $J'$ close to $J$, the transition
is first order, but for $J'$ somewhat less it is second order, and 
for $J'=0.6 J$ the transition temperature is almost exactly the Onsager
temperature. Since the planes have a zero interaction energy in the
ground state and a very small interaction energy at low tempratures,
they behave like almost uncoupled 2D Ising systems. Thus this system
seems to exhibit either a first order transition or a second order
transition with the same sort of ground state, depending on the parameters.

\section{Conclusion}
We have clarified the true nature of the ground states of the 
Ising pyrochlores holmium titanate and dysprosium titanate. We
have pointed out that the ice-like behaviour of dysprosium titanate
seems to arise not from a macroscopic degeneracy of the true ground
state, but from its inaccessibility in practice, and consequently the
tendency of the system to fall into one of a large number of 
slightly excited ice-like states. In holmium titanate, the
ordering temperature is higher than the expected ice-formation
temperature; here, too, the system
gets stuck into excited states, but these are partially ordered states
and the model system shows a clear phase transition. By rigidly
enforcing the ice constraint, we show that this transition exists
independently of the broad ice-state crossover in spin ice, and we
exhibit several models, including the well-known fcc Ising model and
a diamond lattice vertex model, which undergo a similar phase transition.  
Analogous to this vertex model we also exhibit a square lattice vertex
model which has differently ordered ground states depending on 
what interaction parameters we choose, and which we hope to examine
further sometime in the future. 

In addition, we have looked at what happens to dysprosium titanate
when a magnetic field is applied, and compared our conclusions to
available experimental data; we have reproduced earlier experimental
data for a weak field and see similarities in our simulations and the
strong-field data, further confirming our underlying model.  We see
strongly anisotropic behaviour in these systems and predict some
interesting results for possible experiments on single-crystal
samples.  

This work appears as part of the Ph.~D. thesis of R.~S.\cite{thesis}

{\em Note.} After this manuscript was prepared, we learned that the
ordering discussed by us has recently been observed in simulations by
den Hertog {\em et al.}\cite{hertog2}

\end{document}